\documentclass{elsart}

\usepackage{epsfig}

\usepackage{amssymb}

\begin{document}

\begin{frontmatter}



\title{The (2+1)-dimensional charged gravastars}

\author[label1]{Farook Rahaman}\footnote{Corresponding author.}\ead{rahaman@iucaa.ernet.in},
\author[label2]{A. A. Usmani}\ead{anisul@iucaa.ernet.in},
\author[label3]{Saibal Ray}\ead{saibal@iucaa.ernet.in},
\author[label4]{Safiqul Islam}\ead{sofiqul001@yahoo.co.in}

\address[label1]{Department of Mathematics, Jadavpur University,
Kolkata 700 032, West Bengal, India.}
\address[label2]{Department of Physics, Aligarh Muslim University,
Aligarh 202 002, Uttar Pradesh, India}
\address[label3]{Department of Physics, Government College of Engineering and Ceramic
Technology, Kolkata 700 010, West Bengal, India}
\address[label4]{Department of
Mathematics, Jadavpur University, Kolkata 700 032, West Bengal,
India.}

\begin{abstract}
This is a continuation and generalization of our earlier work on
{\it gravastar} in $(2+1)$ anti-de Sitter space-time to $2+1$
dimensional solution of charged gravastar. Morphologically this
gravastar contains three regions namely: (i) charged interior,
(ii) charged shell and (iii) electrovacuum exterior. We have
studied different characteristics in terms of Length and Energy,
Entropy, and Junction conditions of the spherical charged
distribution. It is shown that the present model of charged
gravastar is non-singular and represents itself an alternative of
Black Hole.
\end{abstract}


\end{frontmatter}

\section{Introduction}
Recently the study of {\it gravastars}, the gravitational vacuum
star, have become a subject of considerable interest as it was
proposed as an alternative to black holes. Mazur and Mottola
\cite{Mazur2001,Mazur2004} first proposed a new type of solution
for the endpoint of a gravitational collapse in the form of cold,
dark and compact objects. Therefore, physically this was an
extension of the concept of Bose-Einstein condensate to
gravitational systems. The Mazur-Mottola model
\cite{Mazur2001,Mazur2004} contains an isotropic de Sitter vacuum
in the interior, while the exterior is defined by a Schwarzschild
geometry, separated by a thin shell of stiff matter implying that
the configuration of a gravastar has three different regions with
different equations of state (EOS)
\cite{Visser2004,Bilic2005,Cattoen2005,Carter2005,Lobo2006,DeBenedictis2006,Lobo2007,Horvat2007,Chirenti2007,Rocha2008,Horvat2009,Nandi2009,Turimov2009},
designated by: (I) Interior: $0 \leq r < r_1 $, ~~~$ p = -\rho $;
(II) Shell: $ r_1 < r < r_2 $, ~~~$ p = +\rho $; and (III)
Exterior: $ r_2 < r $, ~~~$ p = \rho =0$. It is argued that the
presence of matter on the thin shell of thickness $r_2-r_1=\delta$
is required to achieve the required stability of systems under
expansion by exerting an inward force to balance the repulsion
from within.

Usmani et al. \cite{Usmani2011} proposed a new model of a
gravastar admitting conformal motion by assuming a charged
interior. Their exterior was defined by a Reissner-Nordstr{\"o}m
line element instead of Schwarzschild's one. Later on Rahaman et
al. \cite{Rahaman2012} designed a neutral spherically symmetric
model of gravastar in $(2+1)$ anti-de Sitter space-time contrary
to the former work where, as usual, space-time was $3+1$
dimensional. The outer region of the Rahaman et al.
\cite{Rahaman2012} model of gravastar corresponds to the exterior
$(2+1)$ anti-de Sitter space-time of BTZ-type black holes as
presented by Ba\~{n}ados, Teitelboim and Zanelli \cite{BTZ1992}.
Therefore, the above two works demand that one should investigate
a $2+1$ dimensional solution for charged gravastar. This is the
motivation of our present investigation.

In favour of inclusion of charge in stellar distribution it has
been argued in the work of Usmani et al. \cite{Usmani2011} that
compact stars tend to assemble a net charge on the surface
\cite{Stettner1973,Whitman1981,Ray2003,Ghezzi2005,Ray2007a}. This
facilitates stability of a fluid sphere by avoiding gravitational
collapse and hence singularity
\cite{Stettner1973,Whitman1981,Felice1995,Sharma2001,Ivanov2002}.
In this connection we would like to mention the interesting
charged model of Horvat \cite{Horvat2009} which represents a
gravastar where the analysis has carried out within Israel's thin
shell formalism and the continuous profile approach.

In the present investigation we have followed the mechanism of
Mazur-Mottola \cite{Mazur2001,Mazur2004} in the framework of
Einstein-Maxwell formalism. To solve the specified equations,
related to the regions designated as Interior, Shell and Exterior,
we have considered appropriate equations of state (EOS), viz.
$\rho = -p$ (dark energy), $ \rho = p$ (stiff fluid) and $ \rho =
p =0$ (dust) respectively. Under these regional classification of
the spherical configuration we have solved the Einstein-Maxwell
field equations in the specific cases. Thereafter we have
characterized the interfaces and shell in terms of Length and
Energy, Entropy. Junction conditions of the spherical charged
distribution are imposed on the different regions by using Lanczos
equations in $(2+1)$ dimensional space-time. The model thus
obtained represents an alternative of black hole in the form of
charged gravastar it is free from singularity.

\section{EINSTEIN-MAXWELL EQUATIONS}
We first consider the line element for the interior space-time of
a static spherically symmetric charged distribution of matter in
$(2+1)$ dimensions in the form \cite{cataldo1992,Martinez2000}
\begin{equation}
ds^2 = -e^{\gamma(r)} dt^2 + e^{\lambda(r)} dr^2 + r^2d\theta^2.
\label{eq1}
\end{equation}
The Hilbert action coupled to electromagnetism is given by
\begin{equation}
I = \int d x^3 \sqrt{-g } \left( \frac{R-2\Lambda}{16 \pi}
-\frac{1}{4} F_a^c F_{bc} + L_{m} \right),\label{eq2}
\end{equation}
where $L_{m}$ is the Lagrangian for matter. The variation with
respect to the metric gives the following self consistent
Einstein-Maxwell equations with cosmological constant $\Lambda$
for a charged perfect fluid distribution:
\begin{equation}
R_{ab} - \frac{1}{2} R g_{ab} +\Lambda g_{ab} = - 8 \pi
(T_{ab}^{PF} +T_{ab}^{EM}).  \label{eq3}
\end{equation}
The explicit forms of the energy momentum tensor (EMT) components
for the matter source (we assumed that the matter distribution at
the interior of the star is perfect fluid type) and
electromagnetic fields are given by:
\begin{equation}
T_{ab}^{PF} = (\rho +p) u_iu_k + p g_{ik},         \label{Eq4}
\end{equation}
\begin{equation}
T_{ab}^{EM} = -\frac{1}{4 \pi } \left( F_a^c F_{bc} -\frac{1}{4}
g_{ab} F_{cd}F^{cd}\right), \label{Eq5}
\end{equation}
where $\rho$, $p$, $u_i$ and $F_{ab}$ are, respectively,
matter-energy density, fluid pressure and velocity three vector of
a fluid element and electromagnetic field. Here, the
electromagnetic field is related to current three vector
\begin{equation}
J^c = \sigma(r) u^c, \label{Eq6}
\end{equation}
as
\begin{equation}
F^{ab}_{;b} = - 4 \pi J^a, \label{Eq7}
\end{equation}
where, $\sigma(r) $ is the proper charge density of the
distribution. In our consideration, the three velocity is assumed
as $u_a = \delta_a^t$ and consequently, the electromagnetic field
tensor can be given as
\begin{equation}
F_{ab}  =  E(r) (\delta_a^t \delta_b^r-\delta_a^r \delta_b^t),
\label{Eq8}
\end{equation}
where $E(r)$ is the electric field.

The Einstein-Maxwell equations with a cosmological constant
($\Lambda < 0$), for the space-time described by the metric
(\ref{eq1}) together with the energy-momentum tensor given in
equations~(3) and (4), yield (rendering $G = c = 1$)
\begin{eqnarray}
\frac{\lambda' e^{-\lambda}}{2r} &=& 8\pi \rho +E^2 +\Lambda,
\label{eq9} \\ \frac{\gamma' e^{-\lambda}}{2r} &=& 8\pi p -E^2
-\Lambda, \label{eq10}  \\ \frac{e^{-\lambda}}{2}\left(\frac{1}{2}
\gamma'^2+\gamma''-\frac{1}{2}\gamma'\lambda'\right) &=& 8\pi p
+E^2 -\Lambda, \label{eq11}\\ \sigma(r) =
\frac{e^{-\frac{\lambda}{2}}}{4 \pi r } (rE)',\label{eq12}
\end{eqnarray}
where a `$\prime$' denotes differentiation with respect to the
radial parameter $r$. The equation (12) can equivalently be
expressed in the form
\begin{equation}
E(r) = \frac{4 \pi }{r} \int_0^r
x\sigma(x)e^{\frac{\lambda(x)}{2}} dx= \frac{q(r)}{r},
\label{Eq13}
\end{equation}
where q(r) is total charge of the sphere under consideration. For
a charged fluid distribution, the generalized
Tolman-Oppenheimer-Volkov (TOV) equation  may be written as
\begin{equation}
\frac{1}{2}\left(\rho + p\right)\gamma' + p' = \frac{1}{8 \pi r^2
} (r^2E^2)', \label{eq14}
\end{equation}
which is the conservation equation in $(2+1)$ dimensions.

We note that the term inside the integral sign in the equation
(13) is $\sigma(r)e^{\frac{\lambda(r)}{2}}$, which is equivalent
to the volume charge density. We will consider the volume charge
density is polynomial function of $r$. Hence we use the condition
\begin{equation}
\sigma(r)e^{\frac{\lambda(r)}{2}} = \sigma_0 r^n, \label{eq15}
\end{equation}
where $n$ is arbitrary constant as polynomial index and the
constant $\sigma_0$ is referred to the central charge density.

By using the latter result in equation (15), one obtains from
equation (13) as
\begin{equation}
E(r) = \frac{4 \pi \sigma_0}{n+2} r^{n+1}, \label{Eq16}
\end{equation}
\begin{equation}
q(r) = \frac{4 \pi \sigma_0}{n+2} r^{n+2}. \label{Eq17}
\end{equation}
Now, we write some consequences of the filed equations and TOV
equation. The equation (9) implies
\begin{equation}
e^{-\lambda(r)} = M(r) - \Lambda r^2,  \label{Eq18}
\end{equation}
where
\begin{equation}
M(r) = C -16 \pi \int_0^r x\rho(x) dx - 2\int_0^r xE^2(x) dx,
\label{Eq19}
\end{equation}
is the active gravitational mass of the spherical distribution.
Since, $e^{\lambda(r)} > 0$ within the charged sphere of radius R
as well as regular at the origin, we demand that $C>0$. Using the
equations (10), (13) and the  above result (18), we finally obtain
the TOV equation as
\begin{equation}
  p' =-\frac{(p+\rho)( 8 \pi p - E^2 - \Lambda )r}{M(r)- \Lambda r^2} +\frac{1}{8 \pi r^2
} (q^2)' \label{eq20}
\end{equation}
Following Mazur and Mottola \cite{Mazur2001,Mazur2004} we consider
a new $2+1$ dimensional charged perfect fluid configuration which
has three different regions with different equations of state:

 [I] Interior: $ 0 \leq  r < R  $,~ $ \rho = -p $;

 [II] Shell: $R \leq r < R+\epsilon$, $ \rho = p$;

 [III] Exterior: $r_2 < r $, $ \rho = p =0$.

Accordingly our configuration is supported by an interior region
with equation of state $p = - \rho$. The shell of our
configuration belongs to the interfaces, $r=R$ and $r=R+\epsilon$,
where $\epsilon$ is the thickness of the shell. This thin shell,
where the metric coefficients are continuous, contains
ultra-relativistic fluid of soft quanta obeying equation of state
$\rho = p$. The outer region of this gravastar corresponds to the
electrovacuum exterior solution popularly known as the charged BTZ
black hole space-time.

\section {Interior region}
Seeking interior solution which is free of any mass-singularity at
the origin, we use the assumption $p=-\rho$ iteratively. We note
that this type of equation of state is available in the literature
and is known as a false vacuum, degenerate vacuum, or
$\rho$-vacuum \cite{Blome1984,Davies1984,Hogan1984,Kaiser1984} and
represents a repulsive pressure. Hence by using the result given
in equation (16), we obtain the following interior solutions:

\begin{equation}
e^\gamma = e^{-\lambda }  =  C- 8 A r^2 R^{2n+2} + \frac{2
B}{(n+2)(2n+2)}r^{2n+4} - \Lambda r^2, \label{eq21}
\end{equation}

\begin{equation}
\rho = -p = \frac{2 \pi \sigma^2_0 (2n+4)}{(2n+2)(n+2)^2}(
R^{2n+2}-r^{2n+2}), \label{eq23}
\end{equation}
where $A = \frac{B(2n+4)}{8(2n+2)}$ and $B= \frac{16\pi^2
\sigma^2_0}{(n+2)^2}$. Here $C$ is an integration constant.

We assume the surface of the charged distribution i.e. shell is
located at $r=R$. Therefore, we have the boundary condition $p(R)
= 0$. Here, we find the active gravitational mass $M(r)$ in the
following form
\begin{equation}
M(r) = \int_0^R 2 \pi r \left( \rho+\frac{E^2}{8 \pi } \right) dr
= \frac{4(2n^2+8n+6)\pi^2\sigma_0^2}{(n+2)^2(2n+2)(2n+4)}
R^{2n+4}\label{eq24}.
\end{equation}
It can be noted from equation (21) that, $C$ being an non-zero
integration constant, the space-time metric thus obtained is free
from any central singularity. It can also be observed via
equations (22) and (23) that the physical parameters, viz.
density, pressure and mass, are dependent on the charge. Therefore
the solutions provide {\it electromagnetic mass} model, such that
for vanishing charge density $\sigma$ all the physical parameters
do not exist
\cite{Lorentz1904,Florides1962,Feynman1964,Cooperstock1978,Tiwari1984,Gautreau1985,Gron1985,Leon1987,Tiwari1991,Ray2004a,Ray2006,Ray2008}.
However, in this connection one interesting point we note that for
the interior region the above mentioned physical parameters in no
way are dependent on the cosmological constant $\Lambda$.

\section{Exterior region of charged gravastar}
The electrovacuum exterior ($p = \rho = 0$) solution corresponds
to a static, charged BTZ black hole is written in the following
form as \cite{cataldo1992}
\begin{equation}
ds ^2 = -\left(-M_0 - \Lambda r^2 - Q^2 \ln r \right) dt^2 +
\left(-M_0 - \Lambda r^2 - Q^2 \ln r\right)^{-1} dr^2 + r^2
d\theta^2.\label{eq25}
\end{equation}

The parameter $M_0$ is the conserved mass associated with
asymptotic invariance under time displacements. This mass is given
by a flux integral through a large circle at space-like infinity.
The parameter $Q$ is total charge of the black hole.

\section{Shell}
We consider thin shell contains ultra-relativistic fluid of soft
quanta obeying  equation of state is $p =\rho$. This assumption is
not new rather known as a stiff fluid and this type of equation of
state which refers to a Zel'dovich Universe have been used by
various authors to study some cosmological
\cite{Zeldovich1972,Carr1975,Madsen1992} as well as astrophysical
phenomena \cite{Braje2002,Linares2004,Wesson1978}.

It is very difficult to solve the field equations within the
non-vacuum region II i.e. within the shell. However, one can
obtain analytic solution within the framework of thin shell limit,
$ 0 < e^{-\lambda} \equiv h << 1$. The advantage of using this
thin shell limit is that in this limit we can set $h$ to be zero
to the leading order. Then the field equations (9) -(11), with
$p=\rho$, may be recast in the forms
\begin{equation}
 h' = - 4 r (\Lambda + E^2),\label{eq26}
\end{equation}

\begin{equation}
\frac{\gamma'}{4} h' =  2 E^2\label{eq27}.
\end{equation}

Integrating (26) immediately to yield
\begin{equation}
h = D - 2\Lambda r^2 -\left[\frac{2 B}{n+2}\right]
r^{2n+4}\label{eq28},
\end{equation}
where $D$ is an integration constant.

 Since, in the
interior of the shell, h takes very very small values, one can
assume $ h \approx o(\epsilon)$ , where $\epsilon$ is the
thickness of the shell. We further assume that thickness of the
shell is very very small. This means $ \epsilon$ is very very
small, $0< \epsilon < < 1$. As a result, we demand that D,
$\Lambda$ and B $< < $  1. In other words, we can take D,
$\Lambda$ and B are all of order $\epsilon$ i.e.  $\approx
o(\epsilon)$.

Employing this value in equation (27), we obtain the other metric
coefficient as
\begin{equation}
e^{\gamma} =  \frac{1}{\left[ \Lambda+ Br^{2n+2}
\right]^{\frac{1}{n+1}}}\label{eq29}.
\end{equation}
Also, using TOV equation (20), one can get
\begin{equation}
8 \pi p = 8 \pi  \rho = \left[\frac{n+2}{n}\right]\left[ \Lambda+
Br^{2n+2} \right]   \label{eq30}.
\end{equation}

Unlike the interior region, we note that the cosmological constant
$\Lambda$ has a definite contribution to the pressure and density
parameters of the shell in an additive manner.

\section{Proper length and Energy}
We assume the interfaces at $r=R$ and $r=R+\epsilon$ describing
the joining surface between two  regions I and  III. Recall the
joining surface  is very thin.   Now, we calculate the proper
thickness between two interfaces i.e. of the shell as:
\begin{eqnarray}
\ell &=& \int _{R}^{R+\epsilon} \sqrt{e^{\lambda} } dr = \int
_{R}^{R+\epsilon} \frac{1}{\sqrt{h(r) }} dr.\label{Eq31}
\end{eqnarray}
Let F is the primitive of $ \frac{1}{\sqrt{h(r) }}$. Then, $
\frac{dF}{dr} =  \frac{1}{\sqrt{h(r) }}$. \\Hence, we get
\begin{eqnarray}
\ell &=& [F] _{R}^{R+\epsilon}.\label{Eq32}
\end{eqnarray}
Using Taylor's theorem, one can  expand $F(R+\epsilon)$ about $R$
and retaining terms up to the first order of $\epsilon$, then,
$F(R+\epsilon) \simeq F(R) + \epsilon F'(R)$ and our $\ell$ would
be $ \ell \approx \epsilon \frac{dF}{dr}|_{r=R}$.

Therefore, the expression for $\ell$ can be written as
\begin{equation}
\ell \approx \frac{\epsilon}{\sqrt{D - 2\Lambda R^2 -\left[\frac{2
B}{n+2}\right] R^{2n+4}}}.\label{eq33}
\end{equation}
The real and positive value of proper length implies  $D
> 2\Lambda R^2 +\left[\frac{2B}{n+2}\right] R^{2n+4}$. However,
it can be noted that the thickness between two interfaces becomes
infinitely large if $D$ takes the value very close to $ 2\Lambda
R^2 +\left[\frac{2 B}{n+2}\right] R^{2n+4}$. On the other hand the
proper length $\ell$ will decrease in the absence of the
cosmological constant $\Lambda$. \\ Obviously, the estimated size
$\ell$ i.e. proper thickness  and coordinate thickness of the
shell are different. Actually, $\ell \approx o(\sqrt{\epsilon})$
as D, $\Lambda$ and B $\approx o(\epsilon)$.

We now calculate the energy $\widetilde{E}$ within the shell only
and we get

$ \widetilde{E}  = 2\pi \int _{R}^{R+\epsilon}  \left[ \rho  +
\frac{E^2}{8 \pi } \right] r dr  \\ = \frac{1}{4} \left[
\frac{(2n+4)\Lambda }{4n}[(R+\epsilon)^2 - R^2] + \frac{(4n+4)B
}{2n(2n+4)}[(R+\epsilon)^{2n+4} - R^{2n+4}]\right]$

Since the thickness $\epsilon$ of the shell is very small i.e.
$\epsilon <<1$, we expand it binomially about $R$ and taking first
order of $\epsilon$, we get
\begin{equation}
\widetilde{E}   \approx \frac{\epsilon}{4} \left[
\frac{(2n+4)\Lambda R }{2n} + \frac{(4n+4)B
}{2n}R^{2n+3}\right]\label{eq34}.
\end{equation}
Obviously here $n \neq 0$. As before, we also notice that the
energy, $\widetilde{E} $ is of order $\epsilon^2$ i.e. $
\widetilde{E} \approx o(\epsilon^2)$.

\section{Entropy}
Adopting concept of Mazur and Mottola \cite{Mazur2001,Mazur2004},
we try to  calculate the entropy of the fluid within the shell:
\begin{equation}
 S = 2\pi\int _{R}^{R+\epsilon} s(r) r \sqrt{e^{\lambda}}dr.\label{eq35}
\end{equation}
Here $s(r)$, the entropy density for the local temperature $T(r)$,
is given by
\begin{equation}
s(r) =  \frac{\alpha^2k_B^2T(r)}{4\pi\hbar^2 } =
\alpha\left(\frac{k_B}{\hbar}\right)\sqrt{\frac{p}{2 \pi
}},\label{eq36}
\end{equation}
where $\alpha^2$ is a dimensionless constant.

Thus the entropy of the fluid within the shell, via the equation
(30), becomes
\begin{eqnarray}
S =  2\pi\int _{R}^{R+\epsilon}
\alpha\left(\frac{k_B}{\hbar}\right) \frac{1}{4 \pi } \frac{
\sqrt{\left[\frac{2n+4}{2n}\right]}\sqrt{\left[ \Lambda r^2+
Br^{2n+4} \right]}dr}{\sqrt{D - 2\Lambda r^2 -\left[\frac{2
B}{n+2}\right] r^{2n+4}}} .\label{eq37}
\end{eqnarray}
Since, thickness of the shell is negligibly small compared to its
position $R$ from the centre of the gravastar (i.e. $\epsilon <<
R$), in the similar way, as we have done above, by expanding the
primitive of the above integrand about $R$ and retaining terms up
to the first order of $\epsilon$, we have  the entropy as
\begin{eqnarray}
S \approx   \alpha\left(\frac{k_B}{\hbar}\right) \frac{\epsilon}{2
} \frac{ \sqrt{\left[\frac{2n+4}{2n}\right]}\sqrt{\left[ \Lambda
R^2+ BR^{2n+4} \right]}}{\sqrt{D - 2\Lambda R^2 -\left[\frac{2
B}{n+2}\right] R^{2n+4}}}.\label{eq38}
\end{eqnarray}

The expression for the entropy shows that the cosmological
constant $\Lambda$ contributes to it a constant part. The
 entropy of the shell is of order $\epsilon$ i.e. $
S \approx o(\epsilon)$.

\section{Junction Condition}
To match the interior region to the exterior electrovacuum
solution at a junction interface S, situated at $r=R$, one needs
to use extrinsic curvature of S. The surface stress energy  and
surface tension of the junction surface S could be determined from
the discontinuity of the extrinsic curvature  of $S$ at $r=R$.
Here the junction surface is a one dimensional ring of matter.
Let, $\eta$ denotes the Riemann normal coordinate at the junction.
We assume $\eta$ be positive in the manifold in region III
described by exterior electrovacuum BTZ spacetime and $\eta$ be
negative in the manifold in region I described by our interior
space-time. In terms of mathematical notations, we have $x^\mu = (
\tau,\phi,\eta) $ and the normal vector components $\xi^\mu = (
0,0,1)$ with the metric $ g_{\eta\eta} =1, ~~ g_{\eta i}~ =~0$ ~~
and~~ $g_{ij} = ~diag~ (~ -1,~r^2 ~)$. The second fundamental
forms associated with the two sides of the shell
\cite{Israel1966,Perry1992,Rahaman2006,Usmani2010,Rahaman2010,Rahaman2011}
are given by
\begin{equation}K^{i\pm}_j =  \frac{1}{2} g^{ik}
\frac{\partial g_{kj}}{\partial \eta}   \mid_{\eta =\pm 0} =
\frac{1}{2}   \frac{\partial r}{\partial \eta} \left|_{r=R}
 ~ g^{ik} \frac{\partial g_{kj}}{\partial r}\right|_{r=R}.
\label{eq39}
\end{equation}
So, the discontinuity in the second fundamental forms is given as
\begin{equation}\kappa _{ij} = K^+_{ij}-K^-_{ij}.
\label{eq40}
 \end{equation}
Now, from Lanczos equation in (2+1) dimensional spacetime,   the
field equations are derived \cite{Perry1992}:
\begin{eqnarray} \sigma &=&  -\frac{1}{8\pi}  \kappa _\phi^\phi,\\
\label{eq41} v &=&  -\frac{1}{8\pi}  \kappa _\tau^\tau,
\label{eq42}
 \end{eqnarray}
where $\sigma$ and $v$ are line energy density and line tension.
Employing relevant information into equations (41) and (42), and
also by setting $r=R$, we obtain
\begin{equation}\sigma =  -\frac{1}{8\pi R}  \left[ \sqrt{ -\Lambda R^2 - M_0 - Q^2 \ln R}
+  \sqrt{ C- 8 \Lambda   R^{2n+4} - \frac{2
B(2n+3)}{(2n+4)(2n+2)}R^{2n+4}}\right], \label{eq43}
\end{equation}
\begin{equation}v =  -\frac{1}{8\pi}  \left[ \frac{-\Lambda R -\frac{Q^2}{2R}}
{\sqrt{ - \Lambda R^2 - M_0- Q^2 \ln R}} +\frac{(- 8 \Lambda
R^{2n+3} -\frac{B(2n+3)}{(2n+2)} R^{2n+3} } {\sqrt{C- 8 \Lambda
R^{2n+4} - \frac{2 B(2n+3)}{(2n+4)(2n+2)}R^{2n+4}}}\right].
\label{eq44}
\end{equation}
Similar to the $(3+1)$ dimensional case the energy density is
negative in the junction shell. It is also noted that the line
tension is negative which implies that there is a line pressure as
opposed to a line tension. The thin shell i.e. region II of our
configuration contains two types of fluid namely, the
ultra-relativistic fluid obeying $p=\rho$ and matter component
with above stress tensor components (43) and (44) that are arisen
due to the discontinuity of second fundamental form  of the
junction interface.

These two non-interacting components of the stress energy tensors
are characterizing features of our non-vacuum region II.

\section{Concluding remarks}
In this letter, we have presented a new model of charged gravastar
in connection to the electrovacuum exterior $(2+1)$ anti-de Sitter
space-time. One of the most interesting features of this model is
that it is free from any singularity and hence represents a
competent candidate in the class of gravastar as an alternative to
black holes \cite{Mazur2001,Mazur2004}.

The solutions obtained here represent {\it electromagnetic mass}
model \cite{Feynman1964}. Historically, Lorentz \cite{Lorentz1904}
proposed his model for extended electron and conjectured that
``there is no other, no `true' or `material' mass,'' and thus
provides only `electromagnetic masses of the electron' whereas
Wheeler \cite{Wheeler1962} and Wilczek \cite{Wilczek1999} pointed
out that electron has a ``mass without mass''. Later on several
works have been carried out by different investigators
\cite{Florides1962,Cooperstock1978,Tiwari1984,Gautreau1985,Gron1985,Leon1987,Tiwari1991,Ray2004a,Ray2006,Ray2008}
under the framework of general relativity.

The cosmological constant $\Lambda$ as proposed by Einstein
\cite{Einstein1917} for stability of his cosmological model has
been adopted here as a purely scalar quantity which has a definite
contribution to the physical parameters with a constant additive
manner. However, due to accelerating phase of the present Universe
\cite{Riess1998,Perlmutter1998}, this erstwhile $\Lambda$ is
now-a-days considered as a dynamical parameter varying with time
in general.  It is therefore a matter of curious issue whether
varying $\Lambda$   has any contribution in formation of the
gravastars. It is to be noted that one should consider $ \Lambda <
0 $ as Cataldo and Cruz \cite{cataldo1992} obtained charged black
hole solution with $ \Lambda < 0 $ that has a horizon. However,
for $ \Lambda > 0 $ a cosmological type horizon exists  with a
naked singularity. Since the  concept of gravastar is to search
configuration which is alternative to black hole, therefore, $
\Lambda < 0 $ is the only possible choice for the gravastar
configuration in three dimensional space time .

\section*{Acknowledgments}
\noindent FR, SR and AAU wish to thank the authorities of the
Inter-University Centre for Astronomy and Astrophysics, Pune,
India for providing the Visiting Associateship under which a part
of this work was carried out. FR is also thankful to  UGC, Govt.
of India, under Research Award Scheme, for providing financial
support.  We are thankful to the anonymous referee for his
valuable comments and constructive suggestions.

\end{document}